\documentclass[10pt, conference]{IEEEtran}
\usepackage[latin9]{inputenc}
\usepackage{url}
\usepackage{color}
\usepackage{soul}
\usepackage{booktabs}
\usepackage{graphicx}
\usepackage{multirow}
\usepackage{citesort}
\usepackage{mathrsfs}
\usepackage{amssymb}
\usepackage{amsbsy}
\usepackage{amsmath}
\usepackage{tcolorbox}
\usepackage{flushend}

\renewenvironment{thebibliography}[1]{%
	\begin{oldthebibliography}{#1}%
		\setlength{\itemsep}{-.3ex}%
	}%
	{%
	\end{oldthebibliography}%
}

\global\long\def\hb{\boldsymbol{h}}

\global\long\def\wb{\boldsymbol{w}}

\global\long\def\alphab{\boldsymbol{\alpha}}

\ifCLASSINFOpdf
\else
\fi
\begin{document}
%

\title{A deep learning model for estimating story points}




%
\author{\IEEEauthorblockN{Morakot Choetkiertikul\IEEEauthorrefmark{1},
Hoa Khanh Dam\IEEEauthorrefmark{1},
Truyen Tran\IEEEauthorrefmark{2},
Trang  Pham\IEEEauthorrefmark{2},
Aditya Ghose\IEEEauthorrefmark{1} and
Tim Menzies \IEEEauthorrefmark{3}}

\IEEEauthorblockA{\IEEEauthorrefmark{1}
	University of Wollongong, Australia\\
	Email: mc650@uowmail.edu.au, hoa@uow.edu.au, aditya@uow.edu.au}

\IEEEauthorblockA{\IEEEauthorrefmark{2}
	Deakin University, Australia\\
	Email: truyen.tran@deakin.edu.au, phtra@deakin.edu.au}

\IEEEauthorblockA{\IEEEauthorrefmark{3}
	North Carolina State University, USA\\
	Email: tim.menzies@gmail.com}
}


\maketitle

\begin{abstract}
Although there has been substantial research in software analytics for effort estimation in traditional software projects, little work has been done for estimation in agile projects, especially estimating user stories or issues. Story points are the most common unit of measure used for estimating the effort involved in implementing a user story or resolving an issue. In this paper, we offer for the \emph{first} time a comprehensive dataset for story points-based estimation that contains 23,313 issues from 16 open source projects. We also propose a prediction model for estimating story points based on a novel combination of two powerful deep learning architectures: long short-term memory and recurrent highway network. Our prediction system is \emph{end-to-end} trainable from raw input data to prediction outcomes without any manual feature engineering. An empirical evaluation demonstrates that our approach consistently outperforms three common effort estimation baselines and two alternatives in both Mean Absolute Error and the Standardized Accuracy.

\end{abstract}


%
\IEEEpeerreviewmaketitle

\section{Introduction}

Effort estimation is an important part of software project management, particularly for planning and monitoring a software project. The accuracy of effort estimation has implications on the outcome of a software project; underestimation may lead to schedule and budget overruns, while overestimation may have a negative impact on organizational competitiveness. Research in software effort estimation dates back several decades and they can generally be divided into model-based and expert-based methods \cite{Menzies2006,Joorgensen2004}. Model-based approaches leverages data from old projects to make predictions about new projects. Expert-based methods rely on human expertise to make such judgements. Most of the existing work (e.g. \cite{Boehm2000a,Sentas,Sentas2005,Kanmani2007,Panda2015,Kanmani2008,Bibi2004,Shepperd1997,Angelis2000,Sarro,Jorgensen2007,Collopy2007,Kocaguneli2012,Kocaguneli2012,Valerdi2011,Chulani1999})
in effort estimation focus on waterfall-like software development. These approaches estimate the effort required for developing a complete software system, relying on a set of features manually designed for characterizing a software project.

In modern agile development settings, software is developed through repeated cycles (iterative) and in smaller parts at a time (incremental), allowing for adaptation to changing requirements at any point during a project's life. A project has a number of \emph{iterations} (e.g. \emph{sprints} in Scrum \cite{Cervone2011}).  An iteration is usually a short (usually 2--4 weeks) period in which the development team designs, implements, tests and delivers a distinct product increment, e.g. a working milestone version or a working release. Each iteration requires the completion of a number of user stories, which are a common way for agile teams to express user requirements. This is a shift from a model where all functionalities are delivered together (in a single delivery) to a model involving a series of incremental deliveries.

There is thus a need to focus on estimating the effort of completing a single user story at a time rather than the entire project. In fact, it has now become a common practice for agile teams to go through each user story and estimate its ``size''. \emph{Story points} are commonly used as a unit of measure for specifying the overall size of a user story \cite{Cohn2005}. Currently, most agile teams heavily rely on experts' subjective assessment (e.g. planning poker, analogy, and expert judgment) to arrive at an estimate. This may lead to inaccuracy and more importantly inconsistencies between estimates. In fact, a recent study \cite{Usman2014} has found that the effort estimates of around half of the agile teams are inaccurate by 25\% or more.


To facilitate research in effort estimation for agile development, we have developed a new dataset for story point effort estimation. This dataset contains 23,313 user stories or issues with ground truth story points. We collected these issues from 16 large open source projects in 9 repositories namely Apache, Appcelerator, DuraSpace, Atlassian, Moodle, Lsstcorp, Mulesoft, Spring, and Talendforge. To the best of our knowledge, this is the \emph{first dataset for story point estimation} where the focus is at the issue/user story level rather than at the project level as in traditional effort estimation datasets.

We also propose a prediction model which supports a team by recommending a story-point estimate for a given user story. Our model learns from the team's previous story point estimates to predict the size of new issues. This prediction system will be used in conjunction with (instead of a replacement for) existing estimation techniques practiced by the team. The key novelty of our approach resides in the combination of two powerful \emph{deep learning} architectures: long short-term memory (LSTM) and recurrent highway network (RHN). LSTM allows us to  model the long-term context in the textual description of an issue, while RHN provides us with a deep representation of that model. We named this approach as Long-Deep Recurrent Neural Network (LD-RNN).

Our LD-RNN model is a fully \emph{end-to-end} system where raw data signals (i.e. words) are passed from input nodes up to the final output node for estimating story points, and the prediction errors are propagated from the output node all the way back to the word layer. LD-RNN automatically learns semantic representations of user story or issue reports, thus liberating the users from manually designing and extracting features.  Our approach consistently outperforms three common baseline estimators and two alternatives in both Mean Absolute Error and the Standardized Accuracy. These claims have also been tested using  a non-parametric Wilcoxon test and Vargha and Delaney's statistic to demonstrate the statistical significance and the effect size.


The remainder of this paper is organized as follows. Section \ref{sect:movitation} provides an example to motivate our work, and the story point dataset is described in Section \ref{section:datasets}. We then present the LD-RNN model and explain how it can be trained in Section \ref{sect:approach} and Section \ref{sect:model-training} respectively. Section  \ref{section:evaluation} reports on the experimental evaluation of our approach. Related work is discussed in Section \ref{section:relatedwork} before we conclude and outline future work in Section \ref{section:conclusion}.

\section{Motivating example}\label{sect:movitation}

\begin{figure}[ht]
	\centering
	\includegraphics[width=\linewidth]{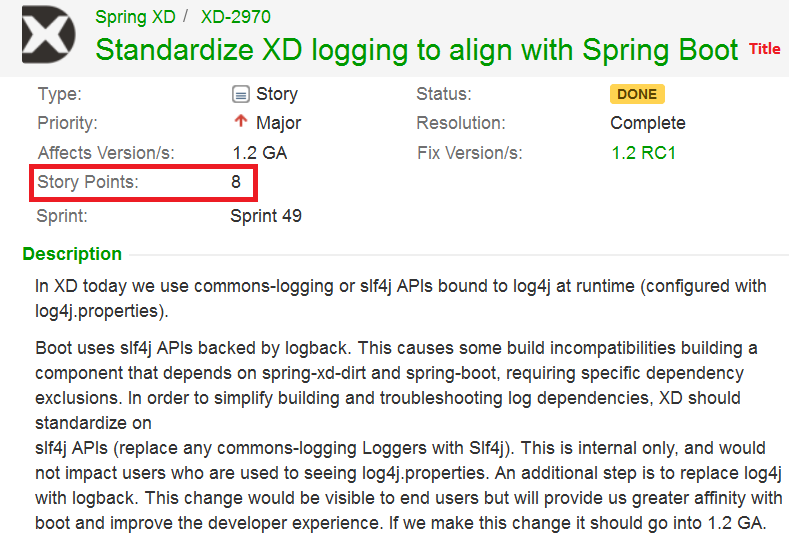}
	\caption{An example of an issue with estimated story points}
	\label{figure:example-story-point}
\end{figure}

When a team estimates with story points, it assigns a point value (i.e. story points) to each user story. A story point estimate reflects the \emph{relative} amount of effort involved in implementing the user story: a user story that is assigned two story points should take twice as much effort as a user story assigned one story point.  Many projects have now adopted this story point estimation approach \cite{Usman2014}. Projects that use issue tracking systems (e.g. JIRA \cite{Atlassian2016a}) record their user stories as \emph{issues}. Figure \ref{figure:example-story-point} shows an example of issue \texttt{XD-2970} in the Spring XD project \cite{Spring2016a} which is recorded in JIRA. An issue typically has a title (e.g. ``Standardize XD logging to align with Spring Boot'') and description. Projects that use JIRA Agile also record story points. For example, the issue in Figure \ref{figure:example-story-point} has 8 story points.

Story points are usually estimated by the whole project team. For example, the widely-used Planning Poker \cite{Grenning2002} method suggests that each team member provides an estimate and a consensus estimate is reached after a few rounds of discussion and (re-)estimation. This practice is different from traditional approaches (e.g. function points) in several aspects. Both story points  and function points are a measure of size. However, function points can be determined by an external estimator based on a standard set of rules (e.g. counting inputs, outputs, and inquiries) that can be applied consistently by any trained practitioner. On the other hand, story points are developed by a specific team based on the team's cumulative knowledge and biases, and thus may not be useful outside the team (e.g. in comparing performance across teams).

Story points are used to compute \emph{velocity}, a measure of a team's rate of progress per iteration. Velocity is the sum of the story-point estimates of the issues that the team resolved during an iteration. For example, if the team resolves four stories each estimated at three story points, their velocity is twelve. Velocity is used for planning and predicting when a software (or a release) should be completed. For example, if the team estimates the next release to include 100 story points and the team's current velocity is 20 points per 2-week iteration, then it would take 5 iterations (or 10 weeks) to complete the project. Hence, it is important that the team is \emph{consistent} in their story point estimates to avoid reducing the predictability in planning and managing their project.

Our proposal enables teams to be {\em consistent} in their estimation of story points. Achieving this consistency is central to effectively leveraging story points for project planning. The machine learner learns from past estimates made by the specific team which it is deployed to assist. The insights that the learner acquires are therefore {\em team-specific}. The intent is not to have the machine learner supplant existing agile estimation practices. The intent, instead, is to deploy the machine learner to {\em complement}  these practices by playing the role of a decision support system. Teams would still meet, discuss user stories and generate estimates as per current practice, but would have the added benefit of access to the insights acquired by the machine learner. Teams would be free to reject the suggestions of the machine learner, as is the case with any decision support system. In every such estimation exercise, the actual estimates generated are recorded as data to be fed to the machine learner, independent of whether these estimates are based on the recommendations of the machine learner or not. This estimation process helps the team not only understand sufficient details about what it will take to to resolve those issues, but also align with their previous estimations.

\section{Story point datasets}
\label{section:datasets}

\begin{table*}[]
	\centering
	\caption{Descriptive statistics of our story point dataset}
	\label{table:dataset}
	\resizebox{6.8in}{!}{%
\begin{tabular}{@{}llcrccccccccr@{}}
	\toprule
	\multicolumn{1}{c}{Repo.} & \multicolumn{1}{c}{Project} & Abb.                 & \multicolumn{1}{c}{\# issues} & min SP               & max SP               & mean SP              & median SP            & mode SP              & var SP               & std SP               & mean TD length       & \multicolumn{1}{c}{LOC} \\ \midrule
	Apache                    & Mesos                       & ME                   & 1,680                         & 1                    & 40                   & 3.09                 & 3                    & 3                    & 5.87                 & 2.42                 & 181.12               & 247,542\textsuperscript{+}                 \\
	& Usergrid                    & UG                   & 482                           & 1                    & 8                    & 2.85                 & 3                    & 3                    & 1.97                 & 1.40                 & 108.60               & 639,110\textsuperscript{+}                 \\
	Appcelerator              & Appcelerator Studio         & AS                   & 2,919                         & 1                    & 40                   & 5.64                 & 5                    & 5                    & 11.07                & 3.33                 & 124.61               & 2,941,856\textsuperscript{\#}               \\
	& Aptana Studio               & AP                   & 829                           & 1                    & 40                   & 8.02                 & 8                    & 8                    & 35.46                & 5.95                 & 124.61               & 6,536,521\textsuperscript{+}               \\
	& Titanium SDK/CLI            & TI                   & 2,251                         & 1                    & 34                   & 6.32                 & 5                    & 5                    & 25.97                & 5.10                 & 205.90               & 882,986\textsuperscript{+}                 \\
	DuraSpace                 & DuraCloud                   & DC                   & 666                           & 1                    & 16                   & 2.13                 & 1                    & 1                    & 4.12                 & 2.03                 & 70.91                & 88,978\textsuperscript{+}                  \\
	Atlassian                      & Bamboo                      & BB                   & 521                           & 1                    & 20                   & 2.42                 & 2                    & 1                    & 4.60                 & 2.14                 & 133.28               & 6,230,465\textsuperscript{\#}               \\
	& Clover                      & CV                   & 384                           & 1                    & 40                   & 4.59                 & 2                    & 1                    & 42.95                & 6.55                 & 124.48               & 890,020\textsuperscript{\#}                 \\
	& JIRA Software               & JI                   & 352                           & 1                    & 20                   & 4.43                 & 3                    & 5                    & 12.35                & 3.51                 & 114.57               & 7,070,022\textsuperscript{\#}               \\
	Moodle                    & Moodle                      & MD                   & 1,166                         & 1                    & 100                  & 15.54                & 8                    & 5                    & 468.53               & 21.65                & 88.86                & 2,976,645\textsuperscript{+}               \\
	Lsstcorp                  & Data Management             & DM                   & 4,667                         & 1                    & 100                  & 9.57                 & 4                    & 1                    & 275.71               & 16.61                & 69.41                & 125,651\textsuperscript{*}                 \\
	Mulesoft                  & Mule                        & MU                   & 889                           & 1                    & 21                   & 5.08                 & 5                    & 5                    & 12.24                & 3.50                 & 81.16                & 589,212\textsuperscript{+}                 \\
	& Mule Studio                 & MS                   & 732                           & 1                    & 34                   & 6.40                 & 5                    & 5                    & 29.01                & 5.39                 & 70.99                & 16,140,452\textsuperscript{\#}                   \\
	Spring                    & Spring XD                   & XD                   & 3,526                         & 1                    & 40                   & 3.70                 & 3                    & 1                    & 10.42                & 3.23                 & 78.47                & 107,916\textsuperscript{+}                 \\
	Talendforge               & Talend Data Quality         & TD                   & 1,381                         & 1                    & 40                   & 5.92                 & 5                    & 8                    & 26.96                & 5.19                 & 104.86               & 1,753,463\textsuperscript{\#}               \\
	& Talend ESB                  & TE                   & 868                           & 1                    & 13                   & 2.16                 & 2                    & 1                    & 2.24                 & 1.50                 & 128.97               & 18,571,052\textsuperscript{\#}                   \\\midrule
	& \multicolumn{1}{c}{Total}   & \multicolumn{1}{l}{} & 23,313                        & \multicolumn{1}{l}{} & \multicolumn{1}{l}{} & \multicolumn{1}{l}{} & \multicolumn{1}{l}{} & \multicolumn{1}{l}{} & \multicolumn{1}{l}{} & \multicolumn{1}{l}{} & \multicolumn{1}{l}{} & \multicolumn{1}{l}{}    \\ \bottomrule
	\multicolumn{13}{c}{SP: story points, TD length: the number of words in the title and description of an issue, LOC: line of code}\\
	\multicolumn{13}{c}{(+: LOC obtained from www.openhub.net, *: LOC from GitHub, and \#: LOC from the reverse engineering)}
\end{tabular}%
	}
\end{table*} 

A number of publicly available datasets (e.g. China, Desharnais, Finnish, Maxwell, and Miyazaki datasets in the PROMISE repository  \cite{Menzies2012a}) have become valuable assets for many research projects in software effort estimation in the recent years. Those datasets however are only suitable for estimating effort at the project level (i.e. estimating effort for developing a complete software system). To the best of our knowledge, there is currently no publicly available dataset for effort estimation at the issue level (i.e. estimating effort for developing a single issue). Thus, we needed to build such a dataset for our study. We have made this dataset {\em publicly available}, both to enable verifiability of our results and also as a service to the research community.

To collect data for our dataset, we looked for issues that were estimated with story points. JIRA is one of the few widely-used issue tracking systems that support agile development (and thus story point estimation) with its JIRA Agile plugin. Hence, we selected a diverse collection of nine major open source repositories that use the JIRA issue tracking system: Apache, Appcelerator, DuraSpace, Atlassian, Moodle, Lsstcorp, MuleSoft, Spring, and Talendforge.
Apache hosts a family of related projects sponsored by the Apache Software Foundation \cite{Apache}.
Appcelerator hosts a number of open source projects that focus on mobile application development \cite{Appcelerator2016}.
DuraSpace contains digital asset management projects \cite{Duraspace2016}.
The Atlassian repository has a number of projects which provide project management systems and collaboration tools \cite{Atlassian2016}.
Moodle is an e-learning platform that allows everyone to join the community in several roles such as user, developer, tester, and QA \cite{Moodle2016}.
Lsstcorp has a number of projects supporting research involving the Large Synoptic Survey Telescope \cite{Lsstcorp2016}.
MuleSoft provides software development tools and platform collaboration tools such as Mule Studio \cite{Mulesoft2016}.
Spring has a number of projects supporting application development frameworks \cite{Spring2016}.
Talendforge is the open source integration software provider for data management solutions such as data integration and master data management \cite{Talendforge2016}.

We then used the Representational State Transfer (REST) API provided by JIRA to query and collected those issue reports. We collected all the issues which were assigned a story point measure from the nine open source repositories up until August 8, 2016. We then extracted the story point, title and description from the collected issue reports. Each repository contains a number of projects, and we chose to include in our dataset only projects that had more than 300 issues with story points. Issues that were assigned a story point of zero (e.g., a non-reproducible bug), as well as issues with a negative, or unrealistically large story point (e.g. greater than 100) were filtered out. Ultimately, about 2.66\% of the collected issues were filtered out in this fashion. In total, our dataset has 23,313 issues with story points from 16 different projects: Apache Mesos (ME), Apache Usergrid (UG), Appcelerator Studio (AS), Aptana Studio (AP), Titanum SDK/CLI (TI), DuraCloud (DC), Bamboo (BB), Clover (CV), JIRA Software (JI), Moodle (MD), Data Management (DM), Mule (MU), Mule Studio (MS), Spring XD (XD), Talend Data Quality (TD), and Talend ESB (TE). Table \ref{table:dataset} summarizes the descriptive statistics of all the projects in terms of the minimum, maximum, mean, median, mode, variance, and standard deviations of story points assigned used and the average length of the title and description of issues in each project. These sixteen projects bring diversity to our dataset in terms of both application domains and project's characteristics. Specifically, they are different in the following aspects: number of observation (from 352 to 4,667 issues), technical characteristics (different programming languages and different application domains), sizes (from 88 KLOC to 18 millions LOC), and team characteristics (different team structures and participants from different regions).

\section{Approach}\label{sect:approach}

Our overall research goal is to build a prediction system that takes as input the title and description of an issue and produces a story-point estimate for the issue. Title and description are required information for any issue tracking system. Although some issue tracking systems (e.g. JIRA) may elicit addition metadata for an issue (e.g. priority, type, affect versions, and fix versions), this information is not always provided at the time that an issues is created. We therefore make a pessimistic assumption here and rely \emph{only} on the issue's title and description. Thus, our prediction system can be used at any time, even when an issue has just been created.

\begin{figure}[ht]
\begin{centering}
\includegraphics[width=2.5in]{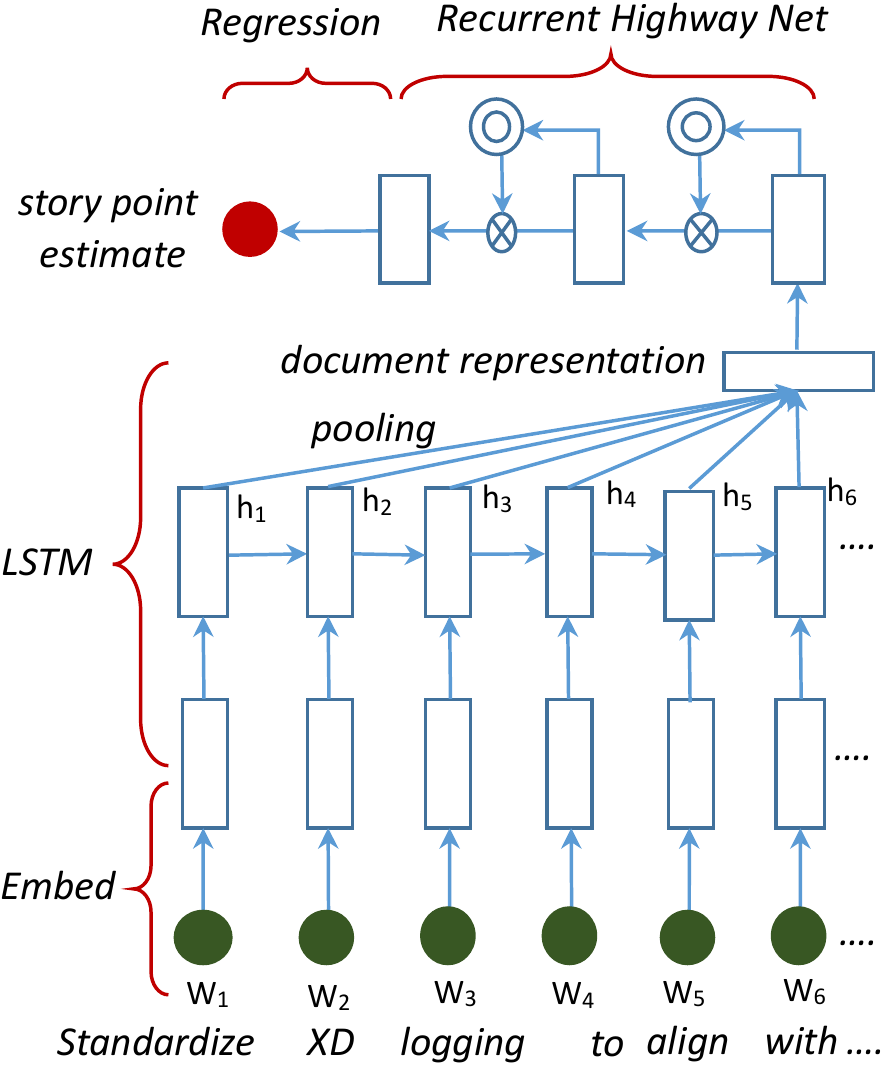}
\par\end{centering}
\caption{Long-Deep Recurrent Neural Net (LD-RNN). The input layer (bottom)
is a sequence of words (represented as filled circles). Words are
first embedded into a continuous space, then fed into the LSTM layer.
The LSTM outputs a sequence of state vectors, which are then pooled
to form a document-level vector. This global vector is then fed into
a Recurrent Highway Net for multiple transformations (See Eq.~(\ref{eq:highway-op})
for detail). Finally, a regressor predicts an outcome (story-point).
\label{fig:Long-Deep-Neural-Net}}
\end{figure}

We combine the title and description of an issue report into a single text document where the title is followed by the description. Our approach computes vector representations for these documents. These representations are then used as features to predict the story points of each issue. It is important to note that these features are \emph{automatically} learned from raw text, hence removing us from manually engineering the features.

Figure \ref{fig:Long-Deep-Neural-Net} shows the Long-Deep Recurrent Neural Network (LD-RNN) that we have designed for the story point prediction system. It is composed of four components arranged sequentially: (i)  word embedding, (ii) document representation using Long Short-Term Memory (LSTM) \cite{hochreiter1997long}, (iii) deep representation using Recurrent Highway Net (RHWN) \cite{pham2016faster}; and (iv) differentiable regression. Given a document which consists of a sequence of words $s=(w_{1},w_{2},...,w_{n})$, e.g. the word sequence \emph{(Standardize, XD, logging, to, align, with, ....)}  in the title and description of issue  \texttt{XD-2970} in Figure \ref{figure:example-story-point}, our LD-RNN can be summarized as follows:
\begin{equation}
y=\text{Regress}\left(\text{RHWN}\left(\text{LSTM}\left(\text{Embed}(s)\right)\right)\right)\label{eq:LD-RNN}
\end{equation}

We model a document's semantics based on the principle of compositionality: the meaning of a document is determined by the meanings of its constituents (e.g. words) and the rules used to combine them (e.g. one word followed by another). Hence, our approach models document representation in two stages. It first converts each word in a document into a fixed-length vector (i.e. word embedding). These word vectors then serve as an input sequence to the Long Short-Term Memory (LSTM) layer which computes a vector representation for the whole document (see Section \ref{sect:docrep} for details).

After that, the document vector is fed into the Recurrent Highway Network (RHWN), which transforms the document vector multiple times (see Section \ref{sect:deeprep} for details), before outputting a final vector which represents the text. The vector serves as input for the \emph{regressor} which predicts the output story-point. While many existing regressors can be employed, we are mainly interested in regressors that are \emph{differentiable} with respect to the training signals and the input vector. In our implementation, we use the simple
\emph{linear regression} that outputs the story-point estimate.

Our entire system is trainable from \emph{end-to-end}: (a) data signals are passed from the words in issue reports to the final output node; and (b) the prediction error is propagated from the output node all the way back to the word layer.

\subsection{Document representation}\label{sect:docrep}

We represent each word as a low dimensional, continuous and real-valued vector, also known as \emph{word embedding}.  Here we maintain a look-up table, which is a word embedding matrix $\mathcal{M} \in \mathbb{R}^{d\times|V|}$ where $d$ is the dimension of word vector and $|V|$ is vocabulary size. These word vectors are pre-trained from corpora of issue reports, which will be described in details in Section \ref{subsec:Pre-training}.

Since an issue document consists of a sequence of words, we model the document by accumulating information from the start to the end of the sequence. A powerful accumulator is a Recurrent Neural Network (RNN) \cite{hochreiter2001gradient}, which can be seen as multiple copies of the same single-hidden-layer network, each passing information to a successor and thus allowing information to be accumulated. While a feedforward neural network maps an input vector into an output vector, an RNN maps a sequence into a sequence. Let $\wb_{1},...,\wb_{n}$ be the input sequence (e.g. words in a sentence). At time step $t$, a standard RNN model reads the input $\wb_{t}$ and the previous state $\hb_{t-1}$ to compute the  state $\hb_{t}$. Due to space limits, we refer the readers to \cite{hochreiter2001gradient} for more technical details of RNN.

While RNNs are theoretically powerful, they are difficult to train
for long sequences \cite{hochreiter2001gradient}, which are often seen in issue reports (e.g. see the description of issue  \texttt{XD-2970} in Figure \ref{figure:example-story-point}).  Hence, our approach employs Long Short-Term Memory (LSTM) \cite{hochreiter1997long,gers2000learning}, a special variant of RNN. The most important element of LSTM is short-term \emph{memory} -- a vector that stores accumulated information over time. The information stored in the memory is refreshed at each time step through partially forgetting old, irrelevant information and accepting fresh new input. However, only some parts of the input will be added to the memory through a selective input gate. Once the memory has been refreshed, an output state will be read from the memory through an output gate. The reading of the new input, writing of the output, and the forgetting are all learnable. LSTM has demonstrated ground-breaking results in many applications such as language models \cite{sundermeyer2012lstm}, speech recognition \cite{graves2013speech} and video analysis \cite{yue2015beyond}. Space limits preclude us from detailing how LSTM works which can be found in its seminal paper \cite{hochreiter1997long}.

After the output state has been computed for every word in the input sequence, a vector representation for the whole document is derived by \emph{pooling} all the vector states. There are multiple ways to perform pooling, but the main requirement is that pooling must be length invariant, that is, pooling is not sensitive to variable length of the document. A simple but often effective pooling method is averaging, which we employed here.

\subsection{Deep representation using Recurrent Highway Network}\label{sect:deeprep}

Given that vector representation of an issue report has been extracted by the LSTM layer, we can use a differentiable regressor for immediate prediction. However, this may be sub-optimal since the network is rather shallow. We have therefore designed a deep representation that performs multiple non-linear
transformations, using the idea from Highway Nets \cite{srivastava2015training}.
A Highway Net is a feedforward neural network that consists of a number of hidden layers, each of which performs a non-linear transformation of the input. Training very deep feedforward networks with many layers is difficult due to two main problems: (i) the number of parameters grows with the number of layers, leading to overfitting; and (ii) stacking many non-linear functions makes it difficult for the information and the gradients to pass through.

Our conception of a Recurrent Highway Network (RHN) addresses the first problem by sharing parameters between layers, i.e. all the hidden layers having the same hidden units (similarly to the notion of a recurrent net). It deals with the second problem by modifying the transformation taking place at a hidden unit to let information
from lower layers pass \emph{linearly} through. Specifically, the hidden state at layer $l$ is defined as:

\begin{equation}
\hb_{l+1}=\alphab_{l}\ast\hb_{l}+\left(\boldsymbol{1}-\alphab_{l}\right)\ast\sigma_{l}\left(\hb_{l}\right)\label{eq:highway-op}
\end{equation}
where $\sigma_{l}$ is a non-linear transform (e.g., a logistic or
a tanh) and $\alphab_{l}=\text{logit}(\hb_{l})$ is a logistic transform
of $\hb_{l}$. Here $\alphab_{l}$ plays the role of a highway gate
that lets information passing from layer $l$ to layer $l+1$ without
loss of information. For example, $\boldsymbol{\alpha}_{l}\rightarrow\boldsymbol{1}$
enables simple copying. This gating scheme is highly effective: while
traditional deep neural nets cannot go beyond several layers, the
Highway Net can have up to a thousand layers \cite{srivastava2015training}.

In previous work \cite{pham2016faster} we found that the operation
in Eq.~(\ref{eq:highway-op}) can be repeated multiple times with
exactly the same set of parameters. In other words, we can create
a very compact version of \emph{Recurrent} Highway Network with only
one set of parameters in $\alphab_{l}$ and $\sigma_{l}$. This clearly
has a great advantage of avoiding overfitting.

\section{Model training}\label{sect:model-training}

\subsection{Training LD-RNN}

We have implemented the LD-RNN model in Python using Theano \cite{Team2016}. To simplify our model, we set the size of the memory cell in an LSTM unit and the size of a recurrent layer in RHWN to be the same as the embedding size. We tuned some important hyper-parameters (e.g. embedding size and the number of hidden layers) by conducting experiments with different values, while for some other hyper-parameters, we used the default values. This will be discussed in more details in the evaluation section.

Recall that the entire network can be reduced to a parameterized function defined in Equation~(\ref{eq:LD-RNN}), which maps sequences of raw words (in issue reports) to story points. Let $\theta$ be the set of all parameters in the model. We define a loss function $L(\theta)$ that measures the quality of a particular set of parameters based on the difference between the predicted story points and the ground truth story points in the training data. A setting of the parameters $\theta$ that produces a prediction for an issue in the training data consistent with its ground truth story points would have a very low loss $L$. Hence, learning is achieved through the optimization process of finding the set of parameters $\theta$ that minimizes the loss function.

Since every component in Equation~(\ref{eq:LD-RNN}) is differentiable, we use the popular stochastic gradient descent to perform optimization: through backpropagation, the model parameters $\theta$ are updated in the opposite direction of the gradient of the loss function $L(\theta)$. In this search, a learning rate $\eta$ is used to control how large of a step we take to reach a (local) minimum. We use RMSprop, an adaptive stochastic gradient method (unpublished note by Geoffrey Hinton), which is known to work best for recurrent models. We tuned RMSprop by partitioning the data into mutually exclusive training, validation, and test sets and running multiple training epoches. Specifically, the training set is used to learn a useful model. After each training epoch, the learned model was evaluated on the validation set and its performance was used to assess against hyperparameters (e.g. learning rate in gradient searches). Note that the validation set was \emph{not} used to learn any of the model's parameters. The best performing model in the validation set was chosen to be evaluated on the test set. We also employed the early stopping strategy, i.e. monitoring the model's performance during the validation phase and stopping when the performance got worse.

To prevent overfitting in our neural network, we have implemented an effective solution called \emph{dropout} in our model \cite{srivastava2014dropout}, where the elements of input and output states are randomly set to zeros during training. During testing, parameter averaging is used. In effect, dropout implicitly trains many models in parallel, and all of them share the same parameter set. The final model parameters represent the average of the parameters across these models. Typically, the dropout rate is set at $0.5$.




An important step prior to optimization is parameter initialization. Typically the parameters are initialized randomly, but our experience shows that a good initialization (through pre-training) helps learning converge faster to good solutions.

\subsection{Pre-training \label{subsec:Pre-training}}

Pre-training is a way to come up with a good parameter initialization \emph{without} using the labels (i.e. ground-truth story points). We pre-train the lower layers of LD-RNN (i.e. embedding and LSTM), which operate at the word level.  Pre-training is effective when the labels are not abundant. During pre-training, we do \emph{not} use the ground-truth story points, but instead leverage two sources of information: the strong predictiveness of natural language, and availability of free texts without labels (e.g. issue reports without story points). The first source comes from the property of languages that the next word can be predicted using previous words, thanks to grammars and common expressions. Thus, at each time step $t$, we can predict the next word $w_{t+1}$ using the state $\hb_{t}$, using the softmax function:
\begin{equation}
P\left(w_{t+1}=k\mid\wb_{1:t}\right)=\frac{\exp\left(U_{k}\hb_{t}\right)}{\sum_{k'}\exp\left(U_{k'}\hb_{t}\right)}\label{eq:lang-softmax}
\end{equation}
where $U_k$ is a free parameter. Essentially we are building a language model, i.e., $P(s)=P\left(\wb_{1:n}\right)$,
which can be factorized using the chain-rule as: $P\left(w_{1}\right)\prod_{t=2}^{n}P\left(w_{t+1}\mid\wb_{1:t}\right)$.

The language model can be learned by optimizing the log-loss $-\log P(s)$. However, the main bottleneck is computational: Equation ~(\ref{eq:lang-softmax}) costs $|V|$ time to evaluate where $|V|$ is the vocabulary size, which can be hundreds of thousands for a big corpus. For that reason, we implemented an approximate but very fast alternative based on Noise-Contrastive Estimation \cite{gutmann2012noise}, which reduces the time to $M\ll |V|$, where $M$ can be as small as 100. We also run multiple epoches against a validation set to choose the best model. We use \emph{perplexity}, a common intrinsic evaluation metric based on the log-loss, as a criterion for choosing the best model and early stopping. A smaller perplexity implies a better language model. The word embedding matrix $\mathcal{M}\in\mathbb{R}^{d\times |V|}$ (which is first randomly initialized) and the initialization for LSTM parameters are learned through this pre-training process.

%







\section{Evaluation}
\label{section:evaluation}

The empirical evaluation we carried out aimed to answer the following research questions:

\begin{itemize}
  \item \textbf{RQ1. Sanity Check}: \emph{Is the proposed approach  suitable for estimating story points?} \\
        This sanity check requires us to compare our LD-RNN prediction model with the three common baseline benchmarks used in the context of effort estimation: Random Guessing, Mean Effort, and Median Effort. Random guessing is a naive benchmark used to assess if an estimation model is useful \cite{Shepperd2012}. Random guessing performs random sampling (with equal probability) over the set of issues with known story points, chooses randomly one issue from the sample, and uses the story point value of that issue as the estimate of the target issue. Random guessing does not use any information associated with the target issue. Thus any useful estimation model should outperform random guessing. Mean and Median Effort estimations are commonly used as baseline benchmarks for effort estimation \cite{Sarro}. They use the mean or median story points of the past issues to estimate the story points of the target issue.

  \item \textbf{RQ2. Benefits of deep representation}: \emph{Does the use of Recurrent Highway Nets provide more accurate and robust estimates than using a traditional regression technique?} \\
      To answer this question, we replaced the Recurrent Highway Net component with a regressor for immediate prediction. Here, we choose a Random Forests (RF) regressor over other baselines (e.g. the one proposed in \cite{Whigham2015}) since ensemble methods like RF, which combine the estimates from multiple estimators, are the most effective method for effort estimation \cite{Kocaguneli2012}. RF achieves a significant improvement over the decision tree approach by generating many classification and regression trees, each of which is built on a random resampling of the data, with a random subset of variables at each node split. Tree predictions are then aggregated through averaging. We then compare the performance of this alternative, namely LSTM+RF, against our LD-RNN model.

  \item \textbf{RQ3. Benefits of LSTM document representation}: \emph{Does the use of LSTM for modeling issue reports provide more accurate results than the traditional Bag-of-Words (BoW) approach?} \\
      The most popular text representation is Bag-of-Words (BoW) \cite{Tirilly2008}, where a text is represented as a vector of word counts. For example, the title and description of issue \texttt{XD-2970} in Figure \ref{figure:example-story-point} would be converted into a sparse binary vector of  vocabulary size, whose elements are mostly zeros, except for those at the positions designated to ``standardize", ``XD", ``logging" and so on. The BoW representation however effectively destroys the sequential nature of text. This question aims to explore whether LSTM with its capability of modeling this sequential structure would improve the story point estimation. To answer this question, we feed two different feature vectors: one learned by LSTM and the other derived from BoW technique to the same Random Forrests regressor, and compare the predictive performance of the former (i.e. LSTM+RF) against that of the latter (i.e. BoW+RF).

  \item \textbf{RQ4. Cross-project estimation}: \emph{Is the proposed approach suitable for cross-project estimation?} \\
     Story point estimation in new projects is often difficult due to lack of training data. One common technique to address this issue is training a model using data from a (source) project and applying it to the new (target) project. Since our approach requires only the title and description of issues in the source and target projects, it is readily applicable to both within-project estimation and cross-project estimation. In practice, story point estimation is however known to be specific to teams and projects. Hence, this question aims to investigate whether our approach is suitable for cross-project estimation. 


\end{itemize}

\subsection{Experimental setting}

We performed experiments on the sixteen projects in our dataset -- see Table \ref{table:dataset} for their details. 
To mimic a real deployment scenario that prediction on a current issue is made by using knowledge from estimations of the past issues, the issues in each project were split into training set (60\% of the issues), development/validation set (i.e. 20\%), and test set  (i.e. 20\%) based on their creation time. The issues in the training set and the validation set were created before the issues in the test set, and the issues in the training set were also created before the issues in the validation set.  

\subsection{Performance measures}

There are a range of measures used in evaluating the accuracy of an effort estimation model. Most of them are based on the Absolute Error, (i.e. $\left | ActualSP - EstimatedSP \right |$). where $AcutalSP$ is the real story points assigned to an issue and $EstimatedSP$ is the outcome given by an estimation model.  Mean of Magnitude of Relative Error (MRE) and Prediction at level l \cite{Conte:1986:SEM:6176}, i.e. Pred($l$),  have also been used in effort estimation.
%
%
However, a number of studies \cite{Foss2003,Kitchenham2001,Korte2008,Port2008} have found that those measures bias towards underestimation and are not stable when comparing effort estimation models. Thus,  the \emph{Mean Absolute Error (MAE)} and the \emph{Standardized Accuracy (SA)} have recently been recommended to compare the performance of effort estimation models \cite{Sarro}. MAE is defined as:

\begin{small}
\[ MAE = \frac{1}{N}\sum_{i=1}^{N}\left | ActualSP_i - EstimatedSP_i \right | \]
\end{small}

where $N$ is the number of issues used for evaluating the performance (i.e. test set), $ActualSP_i$ is the actual story point, and $EstimatedSP_i$ is the estimated story point, for the issue $i$.

SA is based on MAE and it is defined as:

\begin{small}
\[SA = \left ( 1 - \frac{MAE}{MAE_{rguess}} \right )\times  100\]
\end{small}

where  $MAE_{rguess}$ is the MAE of a large number (e.g. 1000 runs) of random guesses. SA measures the comparison against random guessing. Predictive performance can be improved by decreasing MAE or increasing SA.

We assess the story point estimates produced by the estimation models using MAE and SA. To compare the performance of two estimation models, we tested the statistical significance of the absolute errors achieved with the two models using the Wilcoxon Signed Rank Test \cite{Muller1989}. The Wilcoxon test is a safe test since it makes no assumptions about underlying data distributions. The null hypothesis here is: ``the absolute errors provided by an estimation model are significantly less that those provided by another estimation model''. We set the confidence limit at 0.05 (i.e. p $<$ 0.05).

In addition, we also employed a non-parametric effect size measure, the Vargha and Delaney's $\hat{A}_{12}$  statistic \cite{STVR:STVR1486} to assess whether the effect size is interesting. The $\hat{A}_{12}$ measure is chosen since it is agnostic to the underlying distribution of the data, and is suitable for assessing randomized algorithms in software engineering generally \cite{Arcuri2011} and effort estimation in particular \cite{Sarro} . Specifically, given a performance measure (e.g. the Absolute Error from each estimation in our case), the $\hat{A}_{12}$ measures the probability that estimation model $M$ achieves better results (with respect to the performance measure) than estimation model $N$ using the following formula: $\hat{A}_{12} = (r_1/m - (m+1)/2)/n$  where $r_1$ is the rank sum of observations where $M$ achieving better than $N$, and $m$ and $n$ are respectively the number of observations in the samples derived from $M$ and $N$. If the performance of the two models are equivalent, then $\hat{A}_{12} = 0.5$. If $M$ perform better than $N$, then $\hat{A}_{12} > 0.5$ and vice versa. All the measures we have used here are commonly used in evaluating effort estimation models \cite{Sarro,Arcuri2011}.

\subsection{Hyper-parameter settings for training a LD-RNN model}

\begin{figure}[ht]
	\centering
	\includegraphics[width=3.4in]{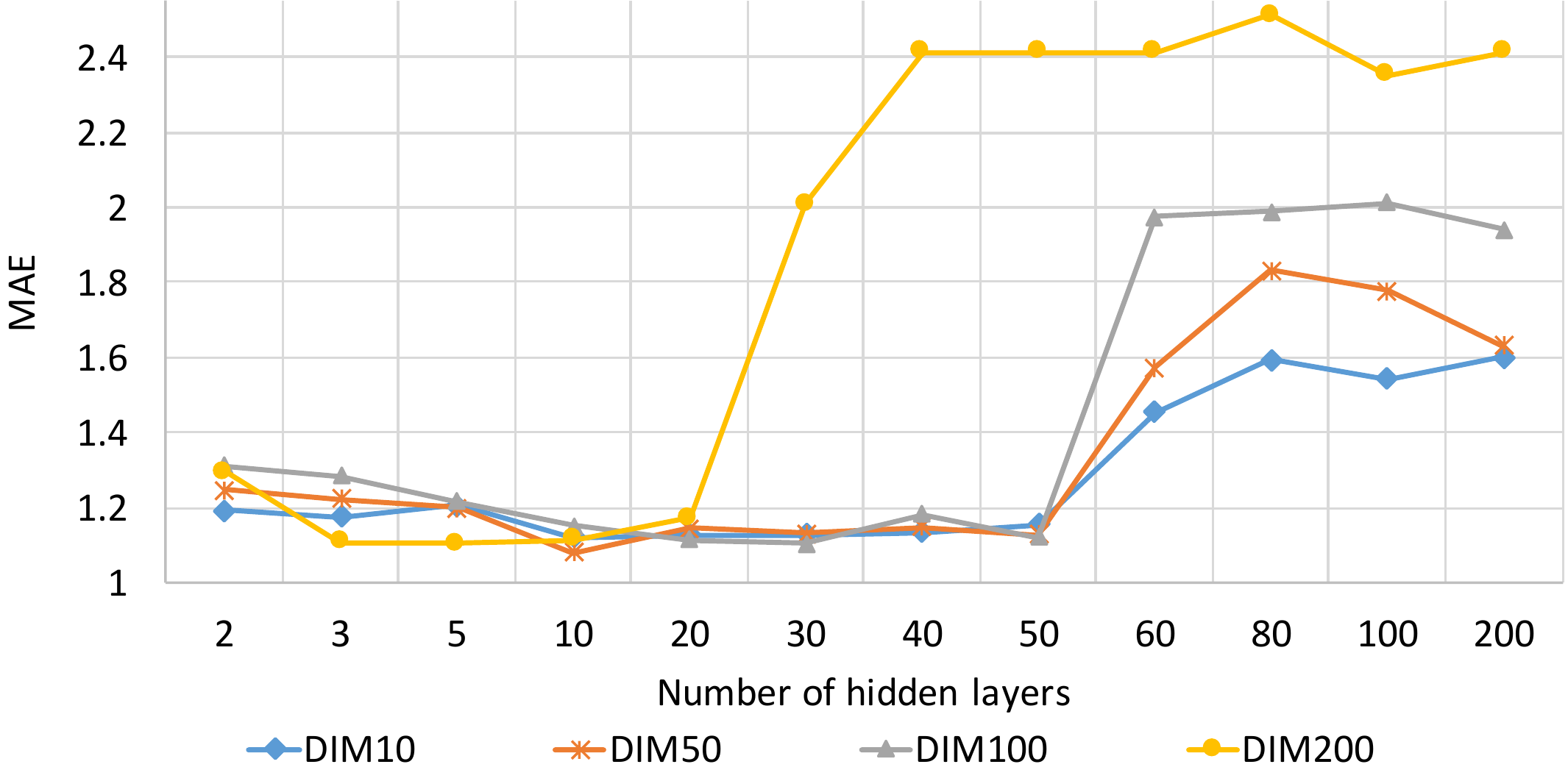}
	\caption{Story point estimation performance with different parameter. }
	\label{figure:diff_dim_hdl}
\end{figure}

We focused on tuning two important hyper-parameters: the number of word embedding dimensions and the number of hidden layers in the recurrent highway net component of our model. To do so, we fixed one parameter and varied the other to observe the MAE performance. We chose to test with four different embedding sizes: 10, 50, 100, and 200, and twelve variations of the number of hidden layers from 2 to 200. This tuning was done using the validation set. Figure \ref{figure:diff_dim_hdl} shows the results from experimenting with Apache Mesos.. As can be seen, the setting where the number of embeddings is 50 and the number of hidden layers is 10 gives the lowest MAE, and thus was chosen.

For both pre-training we ran with 100 epoches and the batch size is 50. The initial learning rate in pre-training was set to $0.02$, adaptation rate was $0.99$, and smoothing factor was $10^{-7}$. For the main LD-RNN model we used 1,000 epoches and the batch size wass set to 100. The initial learning rate in the main model was set to $0.01$, adaptation rate was $0.9$, and smoothing factor was $10^{-6}$. Dropout rates for the RHW and LSTM layers were set to 0.5 and 0.2 respectively.

\subsection{Pre-training}

\begin{figure}[ht]
	\centering
	\includegraphics[width=2.7in]{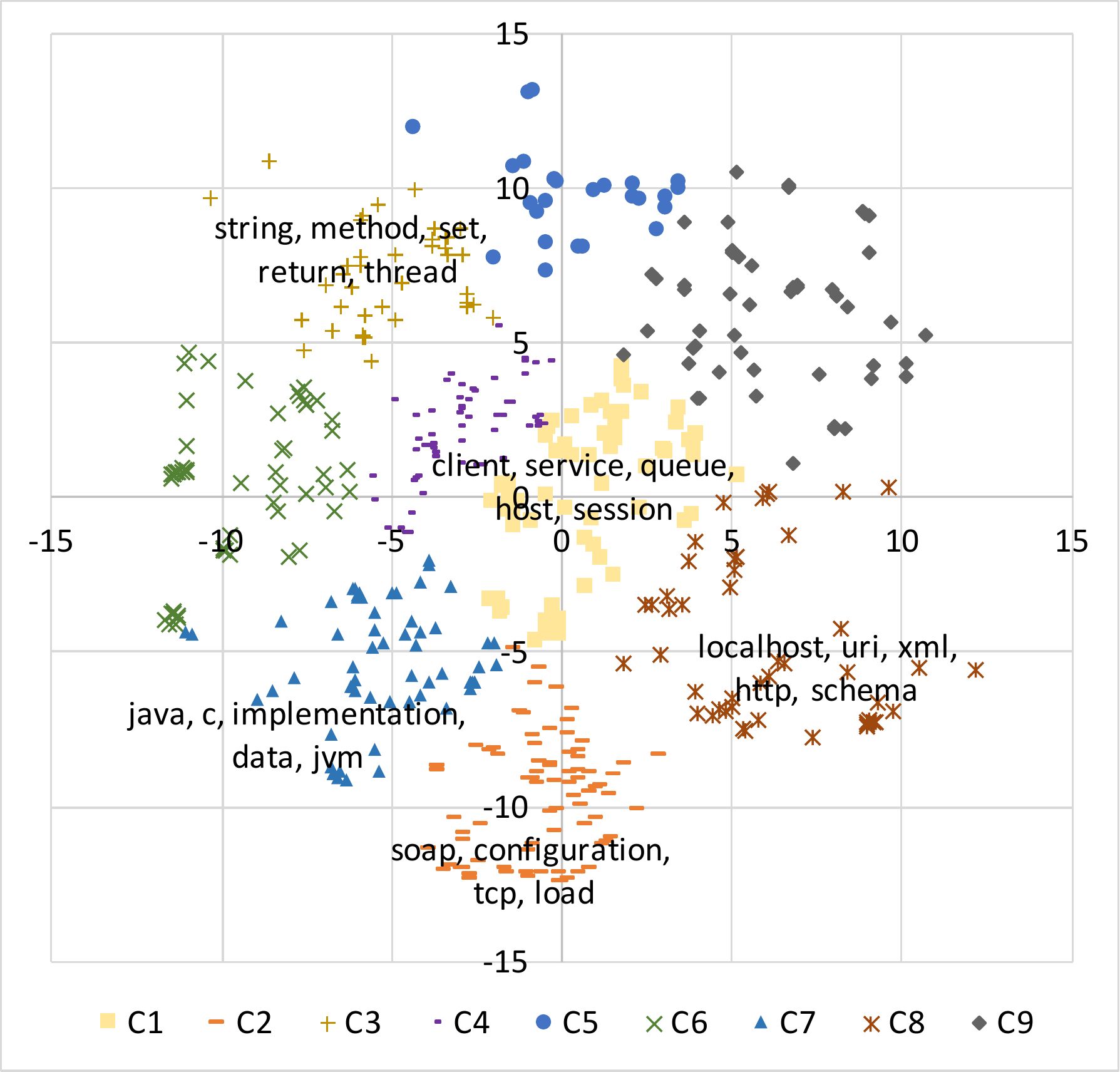}
	\caption{Top-500 word clusters used in the Apache's issue reports}
	\label{figure:word_cluster}
\end{figure}

We used 50,000 issues without story points (i.e. without labels) in each repository for pre-training. Figure \ref{figure:word_cluster} show the top-500 frequent words used in Apache. They are divided into 9 clusters (using K-means clustering) based on their embedding which was learned through the pre-training process. We show here some representative words from some clusters for a brief illustration. Words that are semantically related are grouped in the same cluster. For example, words related to networking like soap, configuration, tcp, and load are in one cluster. This indicates that to some extent, the learned vectors effectively capture the semantic relations between words, which is useful for the story-point estimation task we do later.






\subsection{Results}



We report here the results in answering research questions RQs 1--4.

\begin{table}[ht]
\centering
\caption{Evaluation results (the best results are highlighted in bold). MAE - the lower the better, SA - the higher the better.}
\label{table:overall}
\resizebox{3.2in}{!}{%
\begin{tabular}{@{}clrr|clrr@{}}
\toprule
\multicolumn{1}{c}{Proj} & \multicolumn{1}{c}{Technique} & \multicolumn{1}{c}{MAE} & \multicolumn{1}{c|}{SA} & \multicolumn{1}{c}{Proj} & \multicolumn{1}{c}{Technique} & \multicolumn{1}{c}{MAE} & \multicolumn{1}{c}{SA} \\ \midrule
ME   & LD-RNN    	& \textbf{1.02} & \textbf{59.03} & JI   & LD-RNN    & \textbf{1.38}  & \textbf{59.52}     \\
& LSTM+RF   		& 1.08 			& 57.57 		 &      & LSTM+RF   & 1.71  		 & 49.71 		      \\
& BoW+RF    		& 1.31 			& 48.66 		 &      & BoW+RF    & 2.10  		 & 38.34 		      \\
& Mean      		& 1.64 			& 35.61 		 &      & Mean      & 2.48  		 & 27.06   		      \\
& Median   			& 1.73 			& 32.01 		 &      & Median    & 2.93  		 & 13.88   		      \\ \midrule

UG   & LD-RNN    	& \textbf{1.03} & \textbf{52.66} & MD   & LD-RNN    & \textbf{5.97}  & \textbf{50.29}     \\
& LSTM+RF   		& 1.07 			& 50.70          &		& LSTM+RF   & 9.86  		 & 17.86 		       \\
& BoW+RF    		& 1.19 			& 45.24          &		& BoW+RF    & 10.20 		 & 15.07 		       \\
& Mean      		& 1.48 			& 32.13 		 &		& Mean      & 10.90 		 & 9.16  		       \\
& Median    		& 1.60 			& 26.29 		 &		& Median    & 7.18  		 & 40.16 		      \\ \midrule

AS   & LD-RNN    	& \textbf{1.36} & \textbf{60.26} & DM   & LD-RNN    & \textbf{3.77}  & \textbf{47.87}     \\
& LSTM+RF   		& 1.62 			& 52.38 		 &      & LSTM+RF   & 4.51  		 & 37.71 		      \\
& BoW+RF    		& 1.83 			& 46.34 		 &      & BoW+RF    & 4.78  		 & 33.84 		      \\
& Mean      		& 2.08 			& 39.02 		 &      & Mean      & 5.29  		 & 26.85 		      \\
& Median    		& 1.84 			& 46.17 		 &      & Median    & 4.82  		 & 33.38 		      \\ \midrule

AP   & LD-RNN    	& \textbf{2.71} & \textbf{42.58} & MU   & LD-RNN    & \textbf{2.18}  & \textbf{40.09}     \\
& LSTM+RF   		& 2.97 			& 37.09 		 &		& LSTM+RF   & 2.23  		 & 38.73 		      \\
& BoW+RF    		& 2.96 			& 37.34 		 &      & BoW+RF    & 2.31  		 & 36.64 		      \\
& Mean      		& 3.15 			& 33.30 		 &      & Mean      & 2.59  		 & 28.82 		      \\
& Median    		& 3.71 			& 21.54 		 &      & Median    & 2.69 		  	 & 26.07 		      \\ \midrule

TI   & LD-RNN    	& \textbf{1.97} & \textbf{55.92} & MS   & LD-RNN    & \textbf{3.23}  & \textbf{17.17}     \\
& LSTM+RF   		& 2.32 			& 48.02 		 &      & LSTM+RF   & 3.30  		 & 15.30 		      \\
& BoW+RF    		& 2.58 			& 42.15 		 &      & BoW+RF    & 3.29  		 & 15.58 		      \\
& Mean      		& 3.05 			& 31.59 		 &      & Mean      & 3.34  		 & 14.21 		      \\
& Median    		& 2.47 			& 44.65 		 &      & Median    & 3.30  		 & 15.42 		      \\ \midrule

DC   & LD-RNN    	& \textbf{0.68} & \textbf{69.92} & XD   & LD-RNN    & \textbf{1.63}  & \textbf{46.82}     \\
& LSTM+RF   		& 0.69 			& 69.52 		 &      & LSTM+RF   & 1.81  		 & 40.99 		      \\
& BoW+RF    		& 0.96 			& 57.78 		 &      & BoW+RF    & 1.98  		 & 35.56 		      \\
& Mean      		& 1.30 			& 42.88 		 &      & Mean      & 2.27  		 & 26.00 		      \\
& Median    		& 0.73 			& 68.08 		 &      & Median    & 2.07  		 & 32.55 		      \\ \midrule

BB   & LD-RNN    	& \textbf{0.74} & \textbf{71.24} & TD   & LD-RNN    & \textbf{2.97}  & \textbf{48.28}     \\
& LSTM+RF   		& 1.01 			& 60.95 		 &      & LSTM+RF   & 3.89  		 & 32.14 		      \\
& BoW+RF    		& 1.34 			& 48.06 		 &      & BoW+RF    & 4.49  		 & 21.75 		       \\
& Mean      		& 1.75 			& 32.11 		 &      & Mean      & 4.81  		 & 16.18 		       \\
& Median    		& 1.32 			& 48.72 		 &      & Median    & 3.87  		 & 32.43 		      \\ \midrule

CV   & LD-RNN    	& \textbf{2.11} & \textbf{50.45} & TE   & LD-RNN    & \textbf{0.64}  & \textbf{69.67}     \\
& LSTM+RF   		& 3.08 			& 27.58 		 &      & LSTM+RF   & 0.66  		 & 68.51 		      \\
& BoW+RF    		& 2.98 			& 29.91 		 &      & BoW+RF    & 0.86  		 & 58.89 		      \\
& Mean      		& 3.49 			& 17.84 		 &      & Mean      & 1.14  		 & 45.86 		      \\
& Median    		& 2.84 			& 33.33 		 &      & Median    & 1.16  		 & 44.44 		      \\ \bottomrule
\end{tabular}%
}
\end{table} 

\vspace{0.1cm}
\hspace{-0.4cm}\textbf{RQ1: Sanity check}
\vspace{0.1cm}

The analysis of MAE and SA (see Table  \ref{table:overall}) suggests that the estimations obtained with our approach, LD-RNN, are better than those achieved by using Mean, Median, and Random estimates. LD-RNN consistently outperforms all these three baselines in all sixteen projects. Averaging across all the projects, LD-RNN achieves an accuracy of  2.09 MAE and 52.66 SA, while the best of the baselines achieve only 2.84 MAE and 36.36 SA.


\begin{table}[ht]
	\centering
	\caption{Comparison on the effort estimation benchmarks using Wilcoxon test and $\hat{A}_{12}$ effect size (in brackets)}
	\label{table:LD-RNNvsBenchmark}
	\resizebox{2.9in}{!}{%
		\begin{tabular}{@{}crlrlrl@{}}
			\toprule
			\multicolumn{1}{c}{LD-RNN vs} & \multicolumn{2}{c}{Mean}    & \multicolumn{2}{c}{Median}  & \multicolumn{2}{c}{Random}  \\ \midrule
			ME                            & \textless0.001 & {[}0.70{]} & \textless0.001 & {[}0.74{]} & \textless0.001 & {[}0.76{]} \\
			UG                            & \textless0.001 & {[}0.63{]} & \textless0.001 & {[}0.66{]} & \textless0.001 & {[}0.68{]} \\
			AS                            & \textless0.001 & {[}0.65{]} & \textless0.001 & {[}0.64{]} & \textless0.001 & {[}0.75{]} \\
			AP                            & 0.04           & {[}0.58{]} & \textless0.001 & {[}0.60{]} & \textless0.001 & {[}0.60{]} \\
			TI                            & \textless0.001 & {[}0.72{]} & \textless0.001 & {[}0.64{]} & \textless0.001 & {[}0.73{]} \\
			DC                            & \textless0.001 & {[}0.71{]} & 0.415          & {[}0.54{]} & \textless0.001 & {[}0.81{]} \\
			BB                            & \textless0.001 & {[}0.85{]} & \textless0.001 & {[}0.73{]} & \textless0.001 & {[}0.85{]} \\
			CV                            & \textless0.001 & {[}0.75{]} & \textless0.001 & {[}0.70{]} & \textless0.001 & {[}0.76{]} \\
			JI                            & \textless0.001 & {[}0.76{]} & \textless0.001 & {[}0.79{]} & \textless0.001 & {[}0.69{]} \\
			MD                            & \textless0.001 & {[}0.73{]} & \textless0.001 & {[}0.75{]} & \textless0.001 & {[}0.57{]} \\
			DM                            & \textless0.001 & {[}0.69{]} & \textless0.001 & {[}0.57{]} & \textless0.001 & {[}0.66{]} \\
			MU                            & 0.003          & {[}0.56{]} & \textless0.001 & {[}0.59{]} & \textless0.001 & {[}0.65{]} \\
			MS                            & 0.799          & {[}0.51{]} & 0.842          & {[}0.51{]} & \textless0.001 & {[}0.64{]} \\
			XD                            & \textless0.001 & {[}0.66{]} & \textless0.001 & {[}0.63{]} & \textless0.001 & {[}0.70{]} \\
			TD                            & \textless0.001 & {[}0.77{]} & \textless0.001 & {[}0.63{]} & \textless0.001 & {[}0.67{]} \\
			TE                            & \textless0.001 & {[}0.66{]} & \textless0.001 & {[}0.67{]} & \textless0.001 & {[}0.87{]} \\ \bottomrule
		\end{tabular}%
	}
\end{table}

Table \ref{table:LD-RNNvsBenchmark} shows the results of the Wilcoxon test (together with the corresponding $\hat{A}_{12}$ effect size) to measure the statistical significance and effect size (in brackets) of the improved accuracy achieved by LD-RNN over the baselines: Mean Effort, Median Effort, and Random Guessing.  In 45/48 cases, our LD-RNN significantly outperforms the baselines with effect sizes greater than 0.5.


\begin{tcolorbox}[standard jigsaw,  opacityback=0, left=0pt, right=0pt, top=1pt, boxrule=1pt]
Our approach outperforms the baselines, thus passing the sanity check required by RQ1.
\end{tcolorbox}

\vspace{0.1cm}
\hspace{-0.4cm}\textbf{RQ2: Benefits of deep representation}
\vspace{0.1cm}

The results for the Wilcoxon test to compare our approach of using Recurrent Highway Networks for deep representation of issue reports against using Random Forests coupled with LSTM (i.e. LSTM+RF) is shown in Table  \ref{table:rnn_rf}. The improvement of our approach over LSTM+RF is significant (p $<$ 0.05) with the effect size greater than 0.5 in 13/16 cases. The three projects (DC, MU and MS) where the improvement is not significant have a very small number of issues. It is commonly known that deep highway recurrent networks tend to be significantly effective when we have large datasets.

\begin{table}[ht]
\centering
\caption{Comparison of the Recurrent Highway Net and Random Forests using Wilcoxon test and $\hat{A}_{12}$ effect size (in brackets)}
\label{table:rnn_rf}
\resizebox{2.7in}{!}{%
\begin{tabular}{@{}lc|lc@{}}
\toprule
\multicolumn{1}{c}{Proj} & \multicolumn{1}{c|}{LD-RNN vs LSTM+RF} & \multicolumn{1}{c}{Proj} & \multicolumn{1}{c}{LD-RNN vs LSTM+RF} \\ \midrule
ME   & \textless0.001 {[}0.53{]} & JI   & 0.006 {[}0.64{]}          \\
UG   & 0.004 {[}0.53{]}          & MD   & \textless0.001 {[}0.67{]} \\
AS   & \textless0.001 {[}0.59{]} & DM   & \textless0.001 {[}0.61{]} \\
AP   & \textless0.001 {[}0.60{]} & MU   & 0.846 {[}0.50{]}          \\
TI   & \textless0.001 {[}0.59{]} & MS   & 0.502 {[}0.51{]}          \\
DC   & 0.406 {[}0.54{]}          & XD   & \textless0.001 {[}0.57{]} \\
BB   & \textless0.001 {[}0.64{]} & TD   & \textless0.001 {[}0.62{]} \\
CV   & \textless0.001 {[}0.69{]} & TE   & 0.020 {[}0.53{]}          \\ \bottomrule
\end{tabular}%
}
\end{table} 


When we use MAE and SA as evaluation criteria (see Table  \ref{table:overall}), LD-RNN is still the best approach, consistently outperforming LSTM+RF across all sixteen projects. Averaging across all the projects, LSTM+RF achieves the accuracy of  only 2.61 (versus 2.09 MAE by LD-RNN) and 44.05 (vs.  52.66 SA).

\begin{tcolorbox}[standard jigsaw,  opacityback=0, left=0pt, right=0pt, top=1pt, boxrule=1pt]
The proposed approach of using Recurrent Highway Networks is effective in building a deep representation of issue reports and consequently improving story point estimation.
\end{tcolorbox}

\vspace{0.1cm}
\hspace{-0.4cm}\textbf{RQ3:  Benefits of LSTM document representation}
\vspace{0.1cm}

To study the benefits of using LSTM instead of BoW in representing issue reports, we compared the improved accuracy achieved by Random Forest using the features derived from LSTM against that using the features derived from BoW. For a fair comparison we used Random Forests as the regressor in both settings and the result is reported in Table \ref{table:lstm_bow}. The improvement of LSTM over BoW is significant (p $<$ 0.05) with effect size greater than 0.5 in 13/16 cases. 

\begin{table}[t]
\centering
\caption{Comparison of Random Forest with LSTM and Random Forests with BoW using Wilcoxon test and $\hat{A}_{12}$ effect size (in brackets)}
\label{table:lstm_bow}
\resizebox{2in}{!}{%
\begin{tabular}{@{}cc|cc@{}}
	\toprule
	Proj & LSTM vs BoW               & Proj & LSTM vs BoW               \\ \midrule
ME & \textless0.001 {[}0.58{]} & JI & 0.009 {[}0.60{]}          \\
UG & \textless0.001 {[}0.56{]} & MD & 0.022 {[}0.54{]}          \\
AS & \textless0.001 {[}0.56{]} & DM & \textless0.001 {[}0.54{]} \\
AP & 0.788 {[}0.51{]}          & MU & 0.006 {[}0.53{]}          \\
TI & \textless0.001 {[}0.55{]} & MS & 0.780 {[}0.49{]}          \\
DC & \textless0.001 {[}0.66{]} & XD & \textless0.001 {[}0.54{]} \\
BB & \textless0.001 {[}0.64{]} & TD & \textless0.001 {[}0.60{]} \\
CV & 0.387 {[}0.49{]}          & TE & \textless0.001 {[}0.61{]}\\ \bottomrule
\end{tabular}
}
\end{table} 

LSTM also performs better than BoW with respect to the MAE and SA measures in the same thirteen cases where we used the Wilcoxon test.

\begin{tcolorbox}[standard jigsaw,  opacityback=0, left=0pt, right=0pt, top=1pt, boxrule=1pt]
The proposed LSTM-based approach is effective in automatically learning semantic features representing issue description, which improves story-point estimation.
\end{tcolorbox}

\vspace{0.1cm}
\hspace{-0.4cm}\textbf{RQ4: Cross-project estimation}
\vspace{0.1cm}

We performed  sixteen sets of cross-project estimation experiments to test two settings: (i) within-repository: both the source and target projects (e.g. Apache Mesos and Apache Usergrid) were from the same repository, and pre-training was done using both projects and all other projects in the same repository; and (ii) cross-repository: the source project (e.g. Appcelerator Studio) was in a different repository from the target project Apache Usergrid, and pre-training was done using only the source project.

\begin{table}[ht]
	\centering
	\caption{Evaluation result on cross-project estimation}
	\label{table:crossresult}
	\resizebox{3in}{!}{%
		\begin{tabular}{@{}ccc|ccc|c@{}}
			\toprule
		  \multicolumn{3}{c|}{(i) within-repository} & \multicolumn{3}{c|}{(ii) cross-repository}  \\
			Source       & Target       & MAE  & Source & Target & MAE  & MAE (with-in project) \\ \midrule
			ME           & UG           & 1.16 & AS     & UG     & 1.57 & 1.03                 \\
			UG           & ME           & 1.13 & AS     & ME     & 2.08 & 1.02                 \\
			AS           & AP           & 2.78 & MD     & AP     & 5.37 & 2.71                 \\
			AS           & TI           & 2.06 & MD     & TI     & 6.36 & 1.97                 \\
			AP           & AS           & 3.22 & XD     & AS     & 3.11 & 1.36                 \\
			AP           & TI           & 3.45 & DM     & TI     & 2.67 & 1.97                 \\
			MU           & MS           & 3.14 & UG     & MS     & 4.24 & 3.23                 \\
			MS           & MU           & 2.31 & ME     & MU     & 2.70 & 2.18                 \\\midrule
			\multicolumn{2}{c}{Average} & 2.41 &        &        & 3.51 & 1.93                 \\ \bottomrule
		\end{tabular}%
	}
\end{table} 

Table \ref{table:crossresult} shows the performance of our LD-RNN model for cross-project estimation using the Mean Absolute Error measure.  We used a benchmark of within-project estimation where older issues of the target project were used for training. In all cases, the proposed approach when used for cross-project estimation performed worse than when used for within-project estimation (e.g. on average 24.8\% reduction in performance for within-repository and 81\% for cross-repository). These results confirm a universal understanding \cite{Usman2014} in agile development that story point estimation is specific to teams and projects.

\begin{tcolorbox}[standard jigsaw,  opacityback=0, left=0pt, right=0pt, top=1pt, boxrule=1pt]
Given the specificity of story points to teams and projects, our proposed approach is more effective for within-project estimation.
\end{tcolorbox}


\subsection{Verifiability}

We have created a replication package and will make it publically available soon.
The package contains the full dataset and the source code of our LD-RNN model and the benchmark models (i.e. the baselines, LSTM+RF, and BoW+RF). On this website, we also provide detailed instructions on how to run the code and replicate all the experiments we reported in this paper so that our results can be independently verified.


\subsection{Threats to validity}
\label{section:TOV}

We tried to mitigate threats to construct validity by using real world data from issues recorded in large open source projects. We collected the title and description provided with these issue reports and the actual story points that were assigned to them. To minimize threats to conclusion validity, we carefully selected unbiased error measures and applied a number of statistical tests to verify our assumptions \cite{STVR:STVR1486}. Our study was performed on datasets of different sizes. In addition, we carefully followed recent best practices in evaluating effort estimation models \cite{Arcuri2011,Whigham2015,Shepperd2012} to decrease conclusion instability \cite{Menzies2012b}.  

Training deep neural networks like our LD-RNN system takes a substantial amount of time, and thus we did not have enough time to do cross-fold validation and leave it for future work. One potential research direction is therefore building up a community for sharing pre-trained networks, which can be used for initialization, thus reducing training times (similar to Model Zoo \cite{JiaYangqingandShelhamerEvanandDonahueJeffandKarayevSergeyandLongJonathanandGirshickRossandGuadarramaSergioandDarrell2014}). As the first step towards this direction, we make our pre-trained models publicly available for the research community. Furthermore, our approach assumes that the team stays static over time, which might not be the case in practice. Team changes might affect the set of skills available and consequently story point estimation. Hence, our future work will consider the modeling of team dynamics.

To mitigate threats to external validity, we have considered 23,313 issues from sixteen open source projects, which differ significantly in size, complexity, team of developers, and community. We however acknowledge that our dataset would not be representative of all kinds of software projects, especially in commercial settings (although open source projects and commercial projects are similar in many aspects). One of the key differences between open source and commercial projects that may affect the estimation of story points is the nature of contributors, developers, and project's stakeholders. Further investigation for commercial agile projects is needed.

\section{Related work}
\label{section:relatedwork}


Existing estimation methods can generally be classified into three major groups: expert-based, model-based, and hybrid approaches. Expert-based methods rely on human expertise to make estimations, and are the most popular technique in practice \cite{Joorgensen2004,Jorgensen2009}. Expert-based estimation however tends to require large overheads and the availability of experts each time the estimation needs to be made. Single-expert opinion is also considered less useful on agile projects (than on traditional projects) since estimates are made for user stories or issues which require different skills from more than one person (rather than an expert in one specific task) \cite{Cohn2005}. Hence, group estimation are more encouraged for agile projects such at the planning poker technique \cite{Grenning2002} which is widely used in practice. 

Model-based approaches use data from past projects but they are also varied in terms of building customized models or using fixed models. The well-known construction cost (COCOMO) model \cite{Boehm2000a} is an example of a fixed model where factors and their relationships are already defined. Such estimation models were built based on data from a range of past projects. Hence, they tend to be suitable only for a certain kinds of project that were used to built the model. The customized model building approach requires context-specific data and uses various methods such as regression (e.g. \cite{Sentas,Sentas2005}), Neural Network (e.g. \cite{Kanmani2007,Panda2015}), Fuzzy Logic (e.g. \cite{Kanmani2008}), Bayesian Belief Networks (e.g.\cite{Bibi2004}), analogy-based (e.g. \cite{Shepperd1997,Angelis2000}), and multi-objective evolutionary approaches (e.g.  \cite{Sarro}).  It is however likely that no single method will be the best performer for all project types \cite{Jorgensen2007,Collopy2007,Kocaguneli2012}. Hence, some recent work (e.g. \cite{Kocaguneli2012}) proposes to combine the estimates from multiple estimators. Hybrid approaches (e.g. \cite{Valerdi2011,Chulani1999}) combine expert judgements with the available data -- similarly to the notions of our proposal. 

While most existing work focuses on estimating a whole project, little work has been done in building models specifically for agile projects. Today's agile, dynamic and change-driven projects require different approaches to planning and estimating \cite{Cohn2005}. 
Some recent approaches leverage machine learning techniques to support effort estimation for agile projects. The work in \cite{Abrahamsson:2007} developed an effort prediction model for iterative software development setting using regression models and neural networks. Differing from traditional effort estimation models,  this model is built after each iteration (rather than at the end of a project) to estimate effort for the next iteration. The work in \cite{Hearty2009} built a Bayesian network model for effort prediction in software projects which adhere to the agile Extreme Programming method. Their model however relies on several parameters (e.g. process effectiveness and process improvement) that require learning and extensive fine tuning. 
Bayesian networks are also used in \cite{Perkusich2013} to model dependencies between different factors (e.g. sprint progress and sprint planning quality influence product quality) in Scrum-based software development project in order to detect problems in the project. Our work specifically focuses on estimating issues with story points, which is the key difference from previous work. Previous research (e.g. \cite{Giger2010a,Panjer2007,Bhattacharya2011a,Hooimeijer2007a}) has also been done in predicting the elapsed time for fixing a bug or the delay risk of resolving an issue. However, effort estimation using story points is a more preferable practice in agile development.

\section{Conclusion}
\label{section:conclusion}



In this paper, we have contributed to the research community the \emph{first} dataset for story point estimations, sourcing from 16 large and diverse software projects. We have also proposed a deep learning-based, \emph{fully end-to-end} prediction system for estimating story points, completely liberating the users from manually hand-crafting features. A key novelty of our approach is the combination of two powerful deep learning architectures: Long Short-Term Memory (to learn a vector representation for issue reports) and Recurrent Highway Network (for building a deep representation). The proposed approach has consistently outperformed three common baselines and two alternatives according to our evaluation results.


Our future work would involve expanding our study to commercial software projects and other large open source projects to further evaluate our proposed method. We also consider performing team analytics (e.g. features characterizing a team) to model team changes over time. Furthermore, we will look into experimenting with a sliding window setting to explore incremental learning. In addition, we will also investigate how to best use the issue's metadata (e.g. priority and type) and still maintain the end-to-end nature of our entire model. Our future work also involve comparing our use of the LSTM model against other state-of-the-art models of natural language such as paragraph2vec \cite{Le2014} or Convolutional Neural Network \cite{Kalchbrenner2014}. Finally, we would like to evaluate empirically the impact of our prediction system for story point estimation in practice. This would involve developing the model into a tool (e.g. a JIRA plugin) and then organising trial use in practice. This is an important part of our future work to confirm the ultimate benefits of our approach in general.

\end{document}